\documentclass[twocolumn,showpacs,preprintnumbers,amsmath,amssymb]{revtex4}

\expandafter\let\csname equation*\endcsname\relax
\expandafter\let\csname endequation*\endcsname\relax 

\usepackage{graphicx,amsfonts,amsmath,amssymb,mathrsfs}
\usepackage{graphicx}
\usepackage{dcolumn}
\usepackage{bm}
\usepackage[usenames]{color}

\begin{document}

\newcommand{\x}[1]{{\bf #1}}
\newcommand{\xx}[1]{ {\texttt{ #1}}  }
\newcommand{\exx}{\mathbf{e}_x}
\newcommand{\eyy}{\mathbf{e}_y}
\newcommand{\ezz}{\mathbf{e}_z}
\renewcommand{\r}{\mathbf{r}}
\newcommand{\R}{\mathbf{R}}
\newcommand{\A}{\mathbf{A}}
\newcommand{\B}{\mathbf{B}}
\newcommand{\ta}{\tilde{a}}
\newcommand{\tc}{\widetilde{C}}
\newcommand{\ts}{\widetilde{S}}
\newcommand{\tx}{\tilde{x}}
\newcommand{\ty}{\tilde{y}}
\newcommand{\ti}{\tilde{I}}
\newcommand{\X}{\mathscr{X}}
\newcommand{\Y}{\mathscr{Y}}
\newcommand{\dtX}{\dot{\mathscr{X}}}
\newcommand{\dtY}{\dot{\mathscr{Y}}}
\newcommand{\dX}{\dot{{X}}}
\newcommand{\dY}{\dot{{Y}}}
\newcommand{\dx}{\dot{{x}}}
\newcommand{\dy}{\dot{{y}}}
\newcommand{\df}{\dot{{f}}}
\newcommand{\dg}{\dot{{g}}}
\newcommand{\reg}{\mathscr{R}}
\newcommand{\funz}{\mathscr{F}}
\newcommand{\rr}{\boldsymbol{x}}
\newcommand{\omu}{x_1}
\newcommand{\omd}{x_2}
\newcommand{\omt}{x_3}
\newcommand{\om}{x}
\newcommand{\rotaz}{\mathscr{R}}
\newcommand{\hol}{\mathsf{G}}
\newcommand{\ep}{\varepsilon}
 \newcommand{\ee}{{\mathbf{e}}}
    \newcommand{\xh}{\hat{\mathbf{x}}}
      \newcommand{\xhh}{\hat{\rr}}
 \newcommand{\ii}{{\mathbf{i}}}
\newcommand{\cur}{{\boldsymbol \gamma}}
\newcommand{\curh}{{\hat{\cur}}}
\newcommand{\vv}{{\boldsymbol v}}
\newcommand{\Area}{\Omega}

\title{On the stability  of  quantum holonomic gates}
\author{P Solinas,$^{1,4,5}$, M Sassetti,$^{1, 3}$, P Truini,$^{1,2}$ and N Zangh\`{\i}$^{1,2}$}
\address{
$^1$ Dipartimento di Fisica, Universit\`a di Genova, 
Via Dodecaneso 33, 16146 Genova, Italy}
\address{$^2$ Istituto Nazionale di Fisica Nucleare (Sezione di Genova), 
Via Dodecaneso 33, 16146 Genova, Italy}
\address{$^3$   SPIN-CNR, Via Dodecaneso 33, 16146 Genova, Italy}
\address{$^4$ Low Temperature Laboratory (OVLL), Aalto University School of Science, P.O. Box 13500, 00076 Aalto, Finland}
\address{$^5$ COMP Centre of Excellence, Department of Applied Physics, Aalto University School of Science,
P.O. Box 11000, 00076 Aalto, Finland
}
\email{paolo.solinas@aalto.fi}

\begin{abstract}
{We provide  a unified geometrical description for analyzing the stability  of holonomic quantum gates in the presence of imprecise driving controls (parametric noise).  
We consider the situation in which these fluctuations do not affect the adiabatic evolution but can reduce the logical gate performance. Using the intrinsic geometric properties of the holonomic gates, we show under which conditions on  noise's correlation time  and   strength,   the fluctuations in the driving field cancel out. In this way, we provide theoretical support to previous numerical simulations.  We also briefly comment on the error due to the mismatch between real and nominal time of the period of the driving fields  and show that it can  be reduced by suitably increasing the adiabatic time.
}
\end{abstract}


\maketitle

\section{Introduction}
\label{sec:intro}

Recently, there has been a renewed interest for geometric phases and their application to quantum information   \cite{oreshkov09, oreshkov10,pirkkalainen10, golovach10, solinasPRA10, sjoqvist11, johansson12}  including  several solid state experiments   \cite{leek07, mottonen08, neeley09, pechal11}.
These new results could open the way for the realization of ``holonomic quantum computation'' in which the quantum information is manipulated only by means of geometric operators.
This interesting line of research has been opened by the seminal paper of  Wilczek and Zee \cite{wilczek84} and later on by Zanardi and Rasetti \cite{zanardi99,pachos01}.

Holonomic computation is based on an Hamiltonian $H$ depending adiabatically and periodically on time via a set of time dependent parameters $ \rr(t) \equiv (x_1 (t), \ldots, x_n(t))$.  
Interesting implementation proposals were initially based on atoms and ions \cite{unanyan99,duan01} and later in many other quantum systems \cite{fuentes-guridi02, recati02, faoro03, solinas03,zangh05,kamleitnerPRB11}. 
All these models have an Hamiltonian of the form 
\begin{equation}
\label{projham}
H (\rr) = f(r)\hat{H} (\xhh)\,,
\end{equation} 
where $f$ is a real valued function of $r$, the norm of $\rr\in \mathbb{R}^n$, and $\xhh$ is the unit vector $\rr/r$.
This structure in atomic physics is usually referred as {\it tripod} Hamiltonian.

The logical operation associated with \eqref{projham} depends only on the shadow that the curve $ \rr =\rr(t)$ projects onto the surface of the unit sphere $\mathbb{S}^{n-1}$ (more specifically, on the solid angle spanned by it); moreover,  the dependence of the Hamiltonian on $f(r)$ can be transformed away by suitable (time-dependent) projective transformation in the Hilbert space of the system. Thus the norm $r= r(t)$ does not need not be periodic---only periodicity of $\xhh= \xhh (t)$ is required---and this, in its turn, ensures that any  mismatch between $r(0)$ and $r(T)$ does not affect at all the performance of the gate.

For sake of concreteness, one may think of the specific implementation proposal for quantum dot driven by ultrafast lasers \cite{solinas03}, modulated in amplitude and phase with the quantum information stored in the excitonic degree of freedom.
However, the same formalism and results hold for all the other proposals \cite{unanyan99,duan01,fuentes-guridi02, recati02, faoro03, solinas03,zangh05,kamleitnerPRB11}.

For this purpose, the relevant Hamiltonian is 
\begin{equation}
  H(t) =  \rr{}(t)  \!\cdot \!{} \mathbf{b} 
  \label{basham}
 \end{equation} 
with $\rr{}(t)\in \mathbb{R}^3$ and suitable matrix-valued vector $\mathbf{b}$. More precisely, 
we have the following structure: 
\begin{enumerate}
  \item The   quantum dot has  a  level structure of  three
excited degenerate states $|i\rangle$ ($i=1,2,3$) at energy $\epsilon$,
and  a ground state $|0\rangle$, set for convenience at energy~0.
  \item The system is
driven by time-dependent  laser fields, with frequency in resonance with $\epsilon$, {inducing} transitions between
ground and excited states.  In the interaction representation, the Hamiltonian  governing the dynamics of the system  is  \eqref{basham}, where
$\mathbf{b}$ is the matrix-valued vector with components  
\begin{equation}
b_i= |0\rangle\langle i| + |i\rangle\langle 0|.
\label{eq:b_def}
\end{equation}
  \item The components of the vector  $\rr=\rr{} (t)$ represent the amplitudes of the three laser driving fields which are
  slowly varying functions of time.  Moreover, the motion of the unit vector   $\hat{\rr{}}(t)$
is periodic of 
 period $T$, i.e., \begin{equation} \hat{\rr}(0)= \hat{\rr}(T)\equiv \hat{\rr}_0\,. \label{periodic}
 \end{equation}
\end{enumerate}
Note that we are not assuming periodicity of  $r(t)$.

We refer to the final geometric transformation as the {\it gate operator}    $\hol$, which is
defined as the  adiabatic limit of the evolution  generated by \eqref{basham}, for suitable  logical space $\mathbb{L} \subset \mathbb{C}^4$. (We recall that ``logical space'' stands here for  ``qubit'' or, equivalently, for ``two dimensional complex space $\mathbb{C}^2$''),
\begin{equation}
\left.  \hol =\text{ad--lim } U(T)\right |_{\mathbb{L}}\,, \label{hol}
  \end{equation}
  where $U(t)$ is the evolution operator in $\mathbb{C}^4$ generated by \eqref{basham}, i.e., the solution of ($\hbar=1$ from now on)
  \begin{equation} 
    i\frac{dU}{dt} =   H(t)\, U\,,\quad U(0) = I\,, \label{basequ} 
    \end{equation}
where $I$ is the identity operator.

The main advantage in using the gate operator $\hol$ to manipulate quantum information is that it has an intrinsic robustness under different kinds of errors. In fact, its performance can be optimized in presence of environmental effect  and decoherence \cite{carollo03, wu05, pekola09, solinas10, parodi06, florio06, parodi07, thunstrom05, paladino08}. 
Similar properties and geometric interpretations have been discussed for the Abelian geometric phases for which a full analytical treatment exists \cite{whitney03}. These includes  environmental induced geometric  \cite{whitney05} and non-adiabatic non-Markovian contributions \cite{whitney10}.
Moreover, numerical investigations show  that geometric operators are robust against fluctuations of the driving fields $\rr{}(t)$ \cite{dechiara03,solinas04, zhu05},  so-called parametric noise.  However, no analytic treatment has been provided so far.

Here we provide a first-principle explanation of such a robustness.
We analyze the stability of the gate operator  $\hol$ under different sources of error which can decrease the logical gate performance. We  identify two kind of errors induced by the fluctuations of the driving fields: the error in  switching off   the driving fields and the one accumulated during the evolution.
These errors are supposed to be weak enough to avoid any loss of adiabaticity, i.e., they do not produce any transition between non-degenerate states, but they can  decrease the logical gate performance.
Using the intrinsic geometric properties of the holonomic gates, we show that both errors can be reduced in the adiabatic limit, thus recovering the robustness of the holonomic transformation.

The paper is organized as follows. In Section \ref{agqgates} we set the stage for the study of the stability of $\hol$. In particular, we study the adiabatic regime using the language of scaling limits and introduce the dimensionless adiabatic scaling parameter $\ep$.  In Section \ref{geometry} we highlight the geometrical aspects of the problem, in particular we stress the utility of adopting a geometrical extrinsic point of view in describing the geometrical features of $\hol$. In Section \ref{stab}  we discuss the robustness of the holonomic gate operator in the presence of parametric noise.
In Section \ref{conclusions} we conclude.

\section{Adiabatic regime}\label{agqgates}

In this section we solve    \eqref{basequ} in  three steps.

First, we observe that  $H(t)$ in   (\ref{basham}) is a unitary transformation  of $ H(0)$ multiplied  by a scale factor, i.e.,  
 \begin{equation}
 H(t)= \alpha(t)   R(t)^{-1} H(0) R(t)
 \,.\label{good}
    \end{equation}
 To see how this comes about,  consider the time dependence of the unit vector $\xhh= \xhh (t)$ expressed in terms of spherical coordinates  $\theta= \theta(t)$ and $\phi= \phi(t)$ with respect to 
  an orthonormal basis  $\ii \equiv (\ii_1, \ii_2, \ii_3)$
 in the space $\mathbb{R}^3$ of the driving parameters, i.e., 
  \begin{equation}  \xhh(t) = \sin\theta\cos\phi\, \ii_1+ \sin \theta\sin\phi\,\ii_2+ \cos\theta\,\ii_3\,, \label{xhhx}\end{equation}
 and regard $\mathbb{R}^3$ as embedded in $\mathbb{C}^4$ according to the identifications $\ii_1 =|1\rangle\,,\; \ii_2 =|2\rangle\,,\; \ii_3 =|3\rangle \, ,$
to which one may add, for uniformity of notations, the stipulation  $\ii_0 \equiv|0\rangle$. 

Let  
\begin{equation} 
\mathscr{D}(t)= \begin{bmatrix} 
   \cos\theta(t)\cos\phi(t)  &  \cos \theta(t)\sin\phi(t)      &   -\sin\theta (t)      \\
   -\sin\phi (t) &    \cos\phi (t) &        0\\
  \sin\theta(t)\cos\phi(t)         &      \sin\theta(t)\sin\phi (t)   &   \cos\theta (t)\\ 
\end{bmatrix}
 \label{basicrot}
\end{equation}
in the $(\ii_1, \ii_2, \ii_3)$ basis,  and
\begin{equation} D(t) = \begin{bmatrix} 1 & 0\\
0  & \mathscr{D}(t)   \end{bmatrix} \label{roto} \end{equation}
in the $(\ii_0, \ii_1, \ii_2, \ii_3)$ basis.
Then one may easily check that \eqref{good} is satisfied for \begin{equation} R(t) \equiv  {D}(0)^{-1}  {D}(t) \label{belrot}\end{equation} and   scale factor 
\begin{equation}
\alpha(t) = {r(t)}/{r(0)}\,.
\label{alpha0}
\end{equation}

The second step consists in writing   the equation of motion in the moving frame associated with $R(t)$. This is realized  by  the change of variables    \begin{equation}
V(t)= R(t)U(t) \,, \label{vru}
    \end{equation}
whence by   (\ref{basequ}) and \eqref{good},
 \begin{equation}
i\frac{dV}{dt} = A(t) V +  B(t) V\,, \label{roteq}
    \end{equation}
with
    \begin{equation}
  A(t)\equiv    i \frac{dR(t)}{dt} R(t)^{-1}\quad\text{and}\quad B(t) \equiv \alpha(t) H(0)\,.\label{ab}
    \end{equation}
\medskip

The third step exploits the physical assumption that the external fields are slowly varying functions of time.
Since the energy scale of (\ref{basham}) is determined by $r$ to characterize the evolution we introduce the adiabatic parameter 
\begin{equation}
\varepsilon =\frac{1}{\bar{r}~T}\,,  \label{eq:varepsilon}
\end{equation}
where $\bar{r}$ is an estimate of the size of $r(t)$ (say, the minimum value of $r(t)$ during the drive).  
The parameter $\varepsilon$ must, of course, be small in order to avoid transitions between non-degenerate states.

 For such an adiabatic regime the time dependence of the curve  in $\mathbb{R}^3$  should be better regarded as given by $\rr =\rr(\ep t)$, with the function $\rr (t)$ having  a scale of variation of order one.  The time for the application of the geometric transformation is of order $1/\varepsilon$.
Note that the derivative of $R (t)$  leads to the rescaling $A(t) \to \ep A(\ep t)$, so that the dynamical problem in the adiabatic regime becomes that of solving the rescaled equations of motions for $V(t)$
\begin{equation}
i\frac{dV (t)}{dt} = \ep A(\ep t) V(t) +  B(\ep t) V(t)  \label{rotequ}
    \end{equation}
for  $\ep \ll 1$.

 \label{dia}     
\newcommand{\sfe}{\mathsf{e}}
\newcommand{\ad}{\mathbf{a}}
\newcommand{\ac}{\mathbf{a}^*}

At any given time the Hamiltonian \eqref{basham} can be  diagonalized (the explicit time dependence plays no role in the diagonalization and we shall  omit it  in the notations). 
Consider a vector  $u $ in $\mathbb{C}^4$  written as $ u = u_0\ii_0  + \mathbf{u} $, where $\mathbf{u}= u_1\ii_1 + u_2\ii_2 + u_3\ii_3$ and $\rr= x_1\ii_1 + x_2\ii_2 + x_3\ii_3$. 
From the definitions of $\ii_j$ and $b_j$ in (\ref{eq:b_def}), we obtain  $b_j \ii_k = \delta_{jk} \ii_0 + \delta_{0k} \ii_j$. Immediately from \eqref{basham},  the eigenvalues equation reads
\begin{equation} 
	H   \, [u_0\ii_0  + \mathbf{u}]  = \rr \cdot  \mathbf{u}\, \ii_0 + u_0 \rr  \,,\label{posic}
\end{equation}
from which  eigenvalues and the eigenvectors of $H$ follow: take $\mathbf{u} = \xhh$, 
then from (\ref{posic})
\begin{equation}
 H [u_0\ii_0  + \xhh]  = r[\ii_0 + u_0\xhh]\,. \nonumber
\end{equation}
Thus, $u_0 =1 $ gives the eigenvalue $\lambda_+= r$ and $u_0 =-1 $ gives the eigenvalue $\lambda_-= -r$. The
 corresponding normalized eigenvectors are, respectively,
 \begin{equation}
\ee_{\pm}  =  (1/\sqrt{2}) ( \xhh \pm \ii_0) \\
 \label{eq:bright-states}.
\end{equation}
Finally, for $u_0=0$ and  $\mathbf{u}$ orthogonal to $\xhh$ (as vectors in $\mathbb{R}^3$),  we read from \eqref{posic} that $\lambda_0 =0$ is  a doubly degenerate eigenvalue. 
Thus, we have recovered well known properties of  the tripod Hamiltonian (\ref{basham}): there are two degenerate states, at zero energy, called ``dark states'' and other two, called ``bright states'',  with one excited state ($\ee_+$  in   (\ref{eq:bright-states}))
and one ground state ($ \ee_{-}$ in   (\ref{eq:bright-states})) with energy $r$ and $-r$, respectively.
In the following, we shall denote by $P_{ \beta}  $, $\beta = +1, -1, 0$,  the  spectral projectors of $H(0)$ corresponding respectively to the eigenvalues $\lambda_+ =r(0)$, $\lambda_- =-r(0)$ and $\lambda_0 =0$.   $P_0$ projects onto the plane $\mathbb{L}$ orthogonal to $\xhh (0)$ which soon will be identified with  the logical space  in \eqref{hol}.

Consider now the interaction representation of the time evolution operator $V(t)$, whose dynamics is governed by   \eqref{rotequ},  with respect to free dynamics generated by $B$ 
\begin{equation}V_I(t) =  W (t) V(t)   \,, \label{wv} \end{equation}
with
\begin{equation}W(t)  = e^{i \int_0^t B(\ep t') dt'}= e^{i\frac{h(\ep t)}{\ep} H(0)}=  \sum_\beta  e^{i\frac{h(\ep t)}{\ep} \lambda_\beta} P_{ \beta} \,, \label{ww}\end{equation}
where (see   (\ref{alpha0}))
\begin{equation} h(t) = \int_0^t  \alpha(\tau) d\tau \,. \label{reph}\end{equation} Then  from \eqref{rotequ} it follows that  $V_I(t)$ satisfies 
\begin{equation}
i\frac{dV_I(t)}{dt} = \ep  W (t)A(\ep t)W(t)^{-1} V_I(t)\,, \label{roteque}
    \end{equation}
 or, equivalently
\begin{equation}
V_I(t) =   I+\!\! \int_0^{ \ep t}\!\!\!\!   W ( s/\ep)A(s)W ( s/\ep)^{-1} V_I (s/\ep) ds . \label{rotequet}
\end{equation}

Equation \eqref{rotequet} shows that the  effect of $A$ on the evolution manifests itself only on the adiabatic time scale $t_\ep = t/\ep$. Thus, setting  $V_I^\ep (t) =  V_I (t/\ep)$
one gets
\begin{equation}
V_I^\ep(t) =  I +  \sum_{\beta, \beta'} \int_0^{ \varepsilon t  }  
e^{i\frac{h(s)}{\ep} (\lambda_\beta -\lambda_{\beta'})}
P_\beta A(s)  P_{\beta'} V_I^\ep (s)ds\,. \label{rotequeta0}
    \end{equation}
Since $h(t)$ is  positive and never equal to zero,  standard stationary phase approximation gives $\beta'=\beta$ and 
 \begin{equation}
V_I^\ep(t) =  I +  \sum_{\beta} \int_0^{ \ep t  }  
P_\beta A(s)  P_{\beta} V_I^\ep (s)ds + O(\ep)\,. \label{rotequeta}
    \end{equation}
By multiplying both sides of the above equation   by $P_\alpha$ we see that the evolution separates into autonomous spectral components. In particular,  since $P_0$ projects onto the plane $\mathbb{L}$,    \eqref{rotequeta} defines a non-trivial dynamics  on it. In other words, the operator \begin{equation} G(t) \equiv P_0  V_I (t)P_0  \label{gt} \end{equation}
 evolves autonomously in the adiabatic limit (modulo corrections of order $\ep$) according to the equation
  \begin{equation} i\frac{dG}{dt}  =  \mathcal{A} (t) G \label{gteq} \,,\end{equation}
  where, recalling \eqref{ab}, 
  \begin{equation} \mathcal{A}(t) = i P_0   \frac{dR(t)}{dt} R(t)^{-1} P_0 \,. \label{ag} \end{equation}

\section{Geometry} \label{geometry}

Equation (\ref{ag}) can be solved analytically using a geometric approach.
By defining  the operator-valued vectors $\mathbf{a} = (a_1, a_2, a_3)$ and $\mathbf{a}^* = (a_1^*, a_2^*, a_3^*)$,  having components $a_i= |0\rangle\langle i |$ and $a_i^* = |i \rangle\langle 0|$, $i=1,2,3$, respectively.
The Hamiltonian in   \eqref{basham} can be written as $H(t) =  \mathbf{a} \cdot \rr{}(t) +\mathbf{a}^* \cdot \rr{}(t)  $.
Another   operator-valued vector that it is useful to introduce is  $ {\boldsymbol J} = \mathbf{a}^* \times \mathbf{a} $, 
whose components
\begin{eqnarray}
J_1 &=&  a_2^* a_3 - a_3^* a_2 \nonumber \\
J_2 &=&  a_3^* a_1 - a_1^* a_3\\ 
J_3 &=&  a_1^* a_2 - a_2^* a_1 \,, \nonumber
\label{eq:Jdef}
\end{eqnarray}
are  indeed  the generators of  an $SO(3)$ algebra with commutation relations $[J_i, J_j] = - \epsilon_{ijk} J_k$ and in terms of which  $D(t)$ in \eqref{roto} can be expressed as
 \begin{equation} D(t) = e^{-\theta (t) J_2} e^{-\phi (t) J_3}\label{eqd}
\end{equation}

The  frame $\ee(t) =(\ee_0, \ee_\theta (t),   \ee_\phi (t), \ee_r (t))$, where 
  \begin{eqnarray}
\ee_0 &=& \ii_0 \nonumber \\ 
\ee_r (t)&=& \xhh (t)\nonumber \\
\ee_\theta (t) &=&   \cos\theta(t)\cos\phi(t)\, \ii_1 + \cos \theta(t)\sin\phi(t)\,\ii_2 - \sin\theta(t)\,\ii_3 \label{eeuno} \nonumber  \\
\ee_\phi(t) &=&  -  \sin\phi(t)\, \ii_1+\cos\phi(t)\,\ii_2 \label{eedue}
 \end{eqnarray} 
 is indeed a moving frame adapted  to the surface of the unit sphere $\mathbb{S}^2$ on which $\xhh(t)$ moves in the course of time, i.e.,  $\ee_\theta (t)$ and  $\ee_\phi (t)$ are tangent to $\mathbb{S}^2$, and $\ee_r (t)$ is perpendicular to it.  Then   \eqref{eqd}  is the operator transforming the frame $\ii= (\ii_0, \ii_1, \ii_2, \ii_3)$ into the frame $\ee (t)$,
 \begin{equation} {D}(t)^{-1} \ii_1 = \ee_\theta(t),\quad
{D}(t)^{-1} \ii_2 = \ee_\phi (t),\quad
  {D}(t)^{-1}\ii_3 = \ee_r (t)\,. 
\end{equation}
Note that ${D}(0)^{-1}$ is not the identity and,  in particular, that $ {D}(0)^{-1} \ii_3 = \xhh (0)$, whence from (\ref{belrot}) 
\begin{equation}
\xhh (t) = \ee_r (t)= {D}(t)^{-1}  {D}(0) \xhh (0) = {R}(t)^{-1} \xhh (0) \,.\nonumber
\end{equation}
Then   \eqref{good} is nothing but an expression of the usual duality between action of the operators on vectors and on operators, i.e.,
\begin{equation} \xhh (t) \cdot \mathbf{b}=   \left[ R(t)^{-1} \xhh (0)\right]\cdot \mathbf{b} = \xhh(0) \cdot  \left[R(t)   \mathbf{b} R(t)^{-1} \right]\,, \nonumber \end{equation}
and  $\xhh (0) \cdot \mathbf{b} = D(0)^{-1} b_3 D(0)$.

The unit vectors $ \ee_\theta (t)$ and  $\ee_\phi(t)$  (the  ``dark states'') are a natural basis in the moving degenerate space---the plane orthogonal to $ \ee_r (t) = \xhh (t)$, which is, according to Section \ref{dia}, the degenerate eigenspace of the eigenvalue $0$ of $H(t)$. Let $P_0 (t)$ denote the projector onto such a plane, then $P_0 (0) $ is the projector $P_0$ onto the $\mathbb{L}$ plane. Note that  $P_0$ is just a $D(0)$-rotation of the projector $P (\ii_1  \ii_2)=a_1^* a_1 + a_2^* a_2$  onto the $\ii_1$-$\ii_2$ plane; thus, $P_0 =D(0)^{-1} P (\ii_1  \ii_2) D(0) $.

Putting all the pieces together, from  \eqref{belrot}, \eqref{gteq} and \eqref{ag} we obtain
\begin{equation}
\mathcal{A}(t) =  i D(0)^{-1}P(\ii_1 \ii_2)\frac{dD(t)}{dt}D(t)^{-1}P(\ii_1 \ii_2)D(0)
\nonumber \end{equation}
with  (see   \eqref{eqd}) 
\begin{equation}
\frac{dD(t)}{dt}D(t)^{-1} = -\dot{\theta} J_2 -\dot{\phi} \left( \sin\theta J_1 + \cos\theta J_3 \right) \,.
\nonumber \end{equation}
Using (\ref{eq:Jdef}) and the above definition of $P(\ii_1 \ii_2)$, we can calculate the projection $P(\ii_1 \ii_2) J_k P(\ii_1 \ii_2) =  \delta_{k3 } J_k$ and we are left with
\begin{equation}
\mathcal{A}(t) =  i (\dot{\phi} \cos\theta) D(0)^{-1} J_3 D(0) \,.
\nonumber \end{equation}
Observing \eqref{eqd} at time $t=0$,  and using the rotational properties of $J_k$ we have
\begin{equation}
	D(0)^{-1} J_3 D(0) = {\boldsymbol J}\cdot \xhh(0)
\end{equation}
and, finally, we arrive to a geometric expression
\begin{equation}
\mathcal{A}(t) =  i (\dot{\phi} \cos\theta) {\boldsymbol J}\cdot \xhh(0)\,.\label{aphi}
\end{equation}

We  have at our disposal all the ingredients to determine the geometric operator that is used to manipulate the quantum state.
Now, since $\mathcal{A}(t)$ at different times commutes, equation \eqref{gteq} can be solved by direct exponentiation
\begin{equation} G (t) = \exp \left( i\int_0^t  \cos\theta (t') \dot{\phi}(t') dt' {\boldsymbol J}\cdot \xhh (0)    \right).     \label{finalgt}\end{equation}
One may  recognize that $G(t)$ is a rotation in the $\mathbb{L}$ plane  of an angle given by the integral multiplying ${\boldsymbol J}\cdot~\xhh (0)$ in \eqref{finalgt}. 
At this point starting from   (\ref{gt}) and going back from \eqref{finalgt} to the moving frame by means of equation \eqref{wv}  is immediate:  $W(t)$ is just the identity on $\mathbb{L}$ since the corresponding eigenvalues $\lambda_0$ is zero. To go back to the  laboratory frame, we follow  \eqref{vru} and obtain
 \begin{equation}
 \left.  \text{ad--lim } U(t)\right|_{\mathbb{L}}  =  R (t)^{-1} G(t)\,.
\label{eq:U_op}
 \end{equation}

From the assumption \eqref{periodic} of periodicity of $\xhh (t)$, it follows that $R(T) =I$. Therefore
  \begin{equation} \left. \hol = \text{ad--lim } U(T) \right|_{\mathbb{L}}= G(T) \,. \label{holo}\end{equation}
 If we choose a logical basis directly in  $\mathbb{L}$ we can write $G(t)$ in \eqref{finalgt} as a $2\times 2$ matrix. Thus
 (modulo change of basis in $\mathbb{L}$),
 \begin{equation}
\hol = G(T)  = \begin{bmatrix} \cos \Omega & \sin \Omega\\  -\sin\Omega & \cos\Omega
  \end{bmatrix}    \label{finalgtu} \end{equation}
where  \begin{equation}
\Omega = \int_0^T  \cos\theta (t) \dot{\phi}(t) dt \,.
\label{eq:omega}
    \end{equation}

This is the expected result: the $ G(T) $ operator depends only on the solid angle spanned by $\rr (t)$ on the Hamiltonian parameter space. Note that this result does not rely on the assumption of full periodicity ($\rr(0) =\rr(T)$) made by  Wilczek-Zee  and  Zanardi-Rasetti, although it holds only for a restricted class of Hamiltonians. 

\section{Stability} \label{stab}

We shall now consider the effects of external perturbations on the
system, in order to understand the robustness of quantum evolution in
presence of noise. Several are the sources of undesired energy
exchange due to coupling with impurities, and/or external
environments, or due to imprecise control of the system parameters
during the evolution. In the following we will focus on this
parametric error.  First of all, we observe that the good performance
of the gate $\hol$ relies on the validity of the adiabatic limit, and
since the dimensionless constant $\ep$ is finite, this fact alone
introduces an error of order $\ep$, i.e., $O(\ep)$, which is the order
of magnitude of the off-diagonal terms ($\beta\neq \beta' $) in
\eqref{rotequeta0} which are neglected in the limit $\ep\to 0$.  Then
we shall say that the gate $\hol$ is stable if the perturbations
produce corrections of higher order in $\ep$. Letting aside
corrections not following a power law, we can say that stability is
ensured if the corrections on $\hol$ due to the perturbations are
$O(\ep^r)$, with $r > 1$. To simplify the analysis, we will work with dimensionless
quantities: energy scale in units of $\bar{r}$, so that   \eqref{eq:varepsilon} becomes $\ep = 1/T$.

We consider the error induced by the inaccuracy of the control field.
We start with the case in which the actual curve in the parameter
space is not $\rr(t)$ but instead $ \rr ' (t)= \rr(t) + \delta \rr
(t)$, fulfilling still the periodicity requirement
(\ref{periodic}) on the unit vector $\hat{\rr}'(0)= \hat{\rr}'(T)$.
The error $\delta \rr(t)$ is assumed small with respect to $\rr(t)$ in
the sense of some suitable functional norm.  It should be regarded as
rapidly fluctuating random process whose scale of variation is very
small on the adiabatic scale.

From the above  periodicity it follows that the holonomic operator $U'(T)$, related to $\rr'(t)$  is still of the form
\begin{equation}\left. \text{ad--lim } U'(T) \right|_{\mathbb{L}}= G'(T)= \hol[\rr'(t)] \,. 
\end{equation}
The geometric operator $\hol[\rr'(t)]$ is now a functional of $ \rr ' (t)$ and then the error is accumulated during the whole evolution.

In the following we will evaluate at lowest order in $ \delta \rr (t)$  the error  on the holonomic operator  $U'(T)= \hol[\rr(t)+\delta \rr(t)]$
 induced by the fluctuations of the driving fields along the path. We have 
\begin{equation}
U'(T)= \hol[ \rr(t)+\delta \rr(t)] = \hol [ \rr(t)] +  \delta \hol  + o(\sigma)\,
\label{gprimo}
\end{equation}
where $\sigma$ is a measure of the size of the (mean) variation of $\delta \rr (t)$.
To do that, let us start to rewrite  the solid angle  in   (\ref{eq:omega})  for path $\rr'(t)$
as 
\begin{equation}
\label{solidangle}
\Area[\rr'(t)] = \oint_\curh \cos\theta d\phi =\int_{\Sigma} \sin\theta d\theta d\phi \,,
\end{equation}
where $\curh$ is the shadow on the unit sphere $\mathbb{S}^2$ of the
curve $\cur$ in $\mathbb{R}^3$ given by the parametric equations $\rr'
= \rr' (t)$, and satisfying the conditions \eqref{periodic} of partial
periodicity, $\Sigma$ is the surface on $\mathbb{S}^2$ bounded by
$\curh$, i.e., $\curh = \partial \Sigma$.

Thus, $\Area$ is the area of $\Sigma$, that is, the solid
angle spanned by curve $\cur$ (i.e., the solid angle from which $\cur$
is seen from the origin in $\mathbb{R}^3$). Accordingly, different
unitary geometric transformations and then quantum logical gates can
be constructed traversing different loop in the parameters space.

Since the North Pole $\theta=\phi=0$ is a singularity of spherical coordinates,   one is naturally lead to consider the 1-form on
$\mathbb{S}^2$, locally defined by $\omega = \cos\theta d\phi$ and
extended it to all $\mathbb{S}^2$ in a coordinate independent way.
Accordingly \eqref{solidangle} should be replaced by the Stokes
theorem on $\mathbb{S}^2$ expressed in an ``intrinsic'' geometrical
way (i.e., coordinate independent),
\begin{equation} 
  \Omega = \label{stok}\int_{\partial \Sigma} \omega = \int_{\Sigma}d \omega\,.\end{equation}
This is the approach usually adopted  in holonomic computation \cite{zanardi99,pachos01}. 
To take care of the singularity problem one can rewrite \eqref{solidangle} in terms of the  auxiliary vector field $\A  = \ee_{\phi}( 1- \cos\theta)/(r\sin\theta)$
\begin{equation}
\label{formo} \Area[\rr'(t)]  =   \oint_\curh \A \cdot d\r =  \int\!\!\!\!\int_\Sigma   \B \cdot d\mathbf{S}\,,
\end{equation}
where
\begin{equation} \B = {\boldsymbol \nabla}\times \A =  \frac{1}{r^2} \ee_{r}. \label{magnetic} \end{equation}

We can now rewrite   (\ref{formo}) as  an integral over time
\begin{equation}
\label{omeg}
\Area[\rr'(t)] =  \int_0^T  \A (\xhh'(t)) \cdot \dot{  \xhh}' (t)dt  \,.
\end{equation}
Noticing that the right-hand side is analogous to the Lagrangian of a particle in a magnetic field given by \eqref{magnetic},
we have (see \eqref{finalgt} and \eqref{finalgtu})
\begin{equation}
\delta \hol  =   i [\hol\ {\boldsymbol J}\cdot \xhh (0) ] \delta \Omega\,,
\end{equation}
with  \begin{equation} \delta\Area =   \int_0^T \B (\xhh(t))\times\dot{  \xhh} (t)\cdot \delta \rr(t) \, dt \label{eq:deltaomega} \end{equation} Note, as expected, that radial fluctuations give no contribution to the variation since $\B \times\dot{  \xhh}$ is tangent to the sphere.

We now define the statistical properties of $\delta \rr (t)$ which
describes the parametric noise perturbing the external field.  As
already discussed it should have a scale of variation very small on
the adiabatic scale, yet, if we wish to ensure the validity of the
adiabatic approximation, we should demand, at the same time, that its
scale of variation be sufficiently long on the microscopic scale.  The
simplest possibility to ensure this is to regard the components of
$\delta \rr (t)$ as independent mean-zero stationary Gaussian
processes with ``white-noise'' correlation function, i.e.,
\begin{equation}\label{corre} < \!\delta x_i (t)\delta x_j (t')\! > =
  \delta_{ij} \tau_i \sigma_i^2 \delta (t- t') \,,\end{equation} with
$i, j=1,2,3$ and where $\sigma_i^2$ is a measure of the strength of
the noise components (time-independent, as the processes are
stationary), $\tau_i $s are the correlation times of the noise
components.  Since the Gaussian processes are stationary, we can take
$\sigma_i^2 \equiv <\! \delta x_i(0)^2\!>$.  
    
Again, it is convenient to consider $\tau_i$ and  $\sigma_i$ as functions of $\ep$, e.g., 
\begin{equation}
	\tau_i = \, O(\ep^p)\,,\quad \sigma_i = O(\ep^q) \label{eq:tau_sigma_def}
\end{equation}
with suitable exponents $p>0$ and $q>0$. 
Recalling that the intrinsic error of the adiabatic approximation is $O(\ep)$, we should then  inquire
 whether there is a  range of $p,q$-values for which the mean error
\begin{equation}\label{error}  \Delta\equiv \sqrt{<\!\delta\Area^2\!> - <\!\delta\Area\!>^2}  \end{equation}
is below the $O(\ep)$ upper bound, i.e., $O(\ep^r)$ with $r>1$. Of course,
$\Delta$ would then provide an estimate of the (mean) first order
correction $ \delta \hol $ in \eqref{gprimo}.

Since the fluctuations $\delta x_i$ have zero mean, $<\!\delta\Area\!>=
0$. Moreover,  from \eqref{eq:deltaomega} and  \eqref{corre} it follows
\begin{equation}\label{inte} <\!\delta\Area^2\!> = \sum_{i=1}^3   \tau_i \sigma_i^2 \int_0^T [\B\times \dot{  \xhh} ]_i^2 dt \,.\end{equation}
Recalling \eqref{magnetic},   we find 
\begin{equation}\label{fianlmean}  \Delta^2 = \sum_{i=1}^3   \tau_i \sigma_i^2 \int_0^T \dot{x}_i^2 (t) dt \,.\end{equation}
Since the velocity $\dot{ x}_i$  scales as $1/T$, the integral in the right-hand side of \eqref{fianlmean}  is $ O(T\times 1/T^2)= O(1/T)$, i.e, $O(\ep)$, 
and the mean error is then
\begin{equation}\label{over}  \Delta =  O(\ep^{r})\,, \quad\; r= p/2 + q + 1/2  \,.\end{equation}
From   (\ref{eq:tau_sigma_def}) it follows that with this constraint there are 
many solutions $p/2 + q + 1/2 >1 $, in the desired range of
$p,q$-values, which ensure stability of the gate.  Moreover, from
\eqref{over}, we also read the answer to the questions about the
smallness of $\tau_i$ and largeness of $\sigma_i$, namely, that the
fluctuations can indeed be quite large with respect to $\ep$, provided
that the $\tau_i$s are small on the macroscopic scale (but large on
the microscopic scale). This is the cancellation effect already
discussed in Refs.  \cite{dechiara03, solinas04}.  For example, let
$\ep= 10^{-4}$. Then a noise with correlation time of order, say,
$\tau=10^{-2}$ ($p= 1/2)$ can have fluctuations of order, say,
$\sigma= 10^{-2}$ ($q= 1/2)$ while producing, at the same time, an
error on the gate of order $10^{-5}$ ($ p/2 + q + 1/2= 5/4$), well
within the bound $\ep=10^{-4}$ of the adiabatic regime.

We conclude this section with two observations. The first concerns the
geometric interpretation of solid angle perturbation, with a simple
geometrical formula for the right-hand side of \eqref{inte}. We focus on the
case of noise components $\delta x_i (t)$ with the same statistical
properties, i.e., $\sigma_i=\sigma$ and $\tau_i =\tau$ (a condition
which is indeed quite reasonable from a physical point of view). Since
$\Omega$ is invariant under re-parametrization of time, it is
convenient to use as invariant parameter the arc length $s$ with origin in $\xhh_0$ 
and rewrite \eqref{omeg} as
\begin{equation}
\Area[\rr'(s)] =  \int_0^L  \A (\xhh'(s)) \cdot \dot{  \xhh}' (s) ds\,,
\nonumber
\end{equation}
where $L$ is the length of the curve and   $\dot{  \rr}= d\rr/ds$. Then \eqref{inte} becomes
\begin{equation}
<\!\delta\Area^2\!> =\ell\sigma^2 \sum_{i=1}^3  \int_0^L [\B\times  \dot{  \xhh} ]_i^2 ds \,,
\nonumber
\end{equation}
where  $\ell$ is the correlation length of the noise. 
(Note that  the integral is now $O(1)$, since $L$ is the length of the curve on $\mathbb{S}^2$, and therefore $\ell= O(\ep^{p+1})$). 
But, from   (\ref{magnetic}), $\B = \mathbf{n}$, the normal to the curve lying on $\mathbb{S}^2$, i.e., 
\begin{equation}
<\!\delta\Area^2\!> =   \ell\sigma^2 \int_0^L | \mathbf{n}\times  \dot{  \xhh}| ds \,,
\nonumber
\end{equation}
and one recognizes  \begin{equation} \delta \mathcal{A}\equiv \sigma\int_0^L | \mathbf{n}\times  \dot{  \xhh}| ds \nonumber \end{equation}  
as the area of  a thick boundary along the curve of width  $\sigma$. Thus, $\Delta^2 =\ell\sigma \delta \mathcal{A}\,.$

Let us now turn to the second observation. It concerns how to deal
with the difference between nominal and real value of the period of
the laser.  We recall that in the tripod model two control fields are
turned off at the initial and final configurations
\cite{unanyan99,duan01,fuentes-guridi02, recati02, faoro03,
  solinas03,zangh05,kamleitnerPRB11}. However, in a practical
implementation, there is uncertainty in the real time $T$ in which the
fields are turned off since in general this does not correspond to the nominal time $T_0$, with
$T=T_0 +\Delta T$, being $\Delta T$ a statistical fluctuation time.
This can be due, for example, to latent times or delays in the control fields of the experimental set-up. 
In the parameter space this error corresponds to an evolution that, if referred to the nominal time, has 
$R ^{-1}(T_0) \neq I$, since the periodicity requirement is fulfilled at time $T$, e.g.,  $R ^{-1}(T)=I$.
This means a path non-periodic at the nominal time 
\footnote{Notice that the possibility to have fluctuations at the final time corresponds to an open path in the parameter space.
The more general definition of non-abelian holonomy for open paths and its possible use in quantum computation have been addressed in \cite{kult06}.}.
Thus, from \eqref{eq:U_op}
\begin{equation}
U(T_0) = R^{-1} (T_0) G(T_0) \label{eq:tzero} \,,
\end{equation}
and therefore at the lowest order in $\Delta T$
\begin{equation}
R^{-1} (T_0) = R^{-1} (T-\Delta T) \approx I -\left. \frac{\partial R^{-1}}{\partial T}\right|_T \Delta T \label{eq:rzero}\,.
\end{equation}
For the partial derivative in the right-hand side of the previous equation, one has
\begin{equation}
\left.\frac{\partial R^{-1}}{\partial T}\right|_T = \left.\frac{\partial R^{-1}}{\partial \xhh}\right|_T \cdot \dot{\xhh} (T)
 \,.
\end{equation}
The $ \partial R^{-1}/ {\partial \xhh}$ depends only on geometrical properties of the path and thus is of order one, while the velocity is of order $1/T$ (since we are analyzing the system in the adiabatic regime). Thus the error due to the mismatch between real and nominal time of the period of the laser is of order $1/T\approx 1/T_0$ and can therefore be reduced by suitably increasing the nominal time $T_0$.

\section{Conclusions}   \label{conclusions}

We have provided a unified geometrical description for analyzing the stability of holonomic quantum gates in the presence of parametric noise affecting  their time evolution. We have identified two main critical parameters: the correlation time of the noise and  its strength with respect to the driving field. In this way, we  have recovered  what was already obtained numerically for Abelian \cite{dechiara03} and non-Abelian gates \cite{solinas04}, namely, that even strong fluctuations in the driving field can lead to accurate logical gates if the correlation time is short enough to let the fluctuations cancel out. This is an effect of the geometric dependence of the holonomic operator.
In addition, we have shown that  the error due to the mismatch between real and nominal time of the period of the laser  can  be reduced by suitably increasing the adiabatic time.

\section*{Acknowledgements}
We thank E. De Vito and A. Toigo for  fruitful discussions. N.~Zangh\`\i\ was supported in part by INFN. M. Sassetti acknowledges support from the
EU-FP7 via ITN-2008-234970 NANOCTM and from CNR-SPIN.
This research has been partially supported by the Academy of Finland through its Centres of Excellence Program (project no. 251748).


\section*{References}
\begin{thebibliography}{34}

\bibitem{oreshkov09}  Oreshkov O, Brun T A and Lidar D A 2009 {\it Phys. Rev. Lett.} {\bf 102} 070502
\bibitem{oreshkov10}  Oreshkov O and Calsamiglia J 2010 {\it Phys. Rev. Lett.} {\bf 105} 050503 
\bibitem{pirkkalainen10} Pirkkalainen J M, Solinas P, Pekola J P and M\"ott\"onen M 2010 {\it Phys. Rev. B} {\bf 81} 174506
\bibitem{golovach10} Golovach V N, Borhani M and Loss D 2010 {\it Phys. Rev. A }{\bf 81} 022315
\bibitem{solinasPRA10} Solinas P, Pirkkalainen J M and M\"ott\"onen M 2010 {\it Phys. Rev. A } {\bf 82} 052304
\bibitem{sjoqvist11} Sj\"qvist E, Tong D M , Hessmo B, Johansson M and Singh K 2011 {\it Prerpint} arXiv:1107.5127
\bibitem{johansson12} Johansson M, Sj\"qvist E, Andersson L M, Ericsson M, Hessmo B, Singh K and Tong D M 2012 {\it Prerpint} arXiv:1204.5144

\bibitem{leek07} Leek P J, Fink J M, Blais A, Bianchetti R, G\"oppl M, Gambetta J M, Schuster D I, Frunzio L, Schoelkopf R J and Wallraff A 2007 {\it Science} {\bf 318} 1889
\bibitem{mottonen08} M\"ott\"onen M, Vartiainen J J and Pekola J P 2008 {\it Phys. Rev. Lett.} {\bf 100} 177201
\bibitem{neeley09}  Neeley M {\it et al.} 2009 {\it Science} {\bf 325} 722
\bibitem{pechal11}  Pechal M , Berger S, Abdumalikov A A J, Wallraff A and Filipp S 2012 {\it Phys. Rev. Lett.} {\bf 108} 170401

\bibitem{wilczek84}  Wilczek F and  Zee A 1984 {\it Phys. Rev. Lett.} {\bf 52} 2111
\bibitem{zanardi99}  Zanardi P and Rasetti M 1999 {\it Phys. Lett. A} {\bf 264} 94
\bibitem{pachos01}  Pachos J and Zanardi P 2001 {\it Int. J. Mod. Phys} {\bf B15} 1257

\bibitem{unanyan99} Unanyan R G , Shore B W and Bergmann K 1999 {\it Phys. Rev. A } {\bf 59} 2910
\bibitem{duan01} Duan L M, Cirac J I and Zoller P 2001 {\it Science} {\bf 292} 1695

\bibitem{fuentes-guridi02} Fuentes-Guridi I, Pachos J, Bose S, Vedral V and Choi S 2002 {\it Phys. Rev. A } {\bf 66} 022102
\bibitem{recati02} Recati A, Calarco T, Zanardi P, Cirac J I and Zoller P 2002 {\it Phys. Rev. A }{\bf 66} 032309
\bibitem{faoro03} Faoro L, Siewert J and Fazio R 2003 {\it Phys. Rev. Lett.} {\bf 90} 028301
\bibitem{solinas03} Solinas P, Zanardi P, Zangh\`i N and Rossi F 2003 {\it Phys. Rev. B} {\bf 67} 121307(R)
\bibitem{zangh05} Zhang P, Wang Z D, Sun J D and Sun C P 2005 {\it Phys. Rev. A }{\bf 71} 042301
\bibitem{kamleitnerPRB11} Kamleitner I, Solinas P, M\"uller C, Shnirman A, M\"ott\"onen M 2011 {\it Phys. Rev. B} {\bf 83} 214518
\bibitem{carollo03} Carollo A, Fuentes-Guridi I, Santos M F, and Vedral V 2003 {\it Phys. Rev. Lett.} {\bf 90} 160402
\bibitem{wu05}  Wu L A, Zanardi P and Lidar D A 2005 {\it Phys. Rev. Lett.} {\bf 95} 130501
\bibitem{pekola09} Pekola J P, Brosco V, M\"ott\"onen M, Solinas P and Shnirman A 2010 {\it Phys. Rev. Lett.} {\bf 105} 030401
\bibitem{solinas10} Solinas P, M\"ott\"onen M, Salmilehto J and Pekola J P 2010 {\it Phys. Rev. B} {\bf  82} 134517

\bibitem{parodi06}  Parodi D, Sassetti M, Solinas P, Zanardi P and Zangh\`{\i} N 2006 {\it Phys. Rev. A }{\bf 73} 052304
\bibitem{florio06} Florio G, Facchi P, Fazio R, Giovannetti V and Pascazio S 2006 {\it Phys. Rev. A} {\bf 73} 022327
\bibitem{parodi07} Parodi D, Sassetti M, Solinas P  and Zangh\`{\i} N 2007 {\it Phys. Rev. A} {\bf 76} 012337
\bibitem{thunstrom05}Thunstr\"om P, Aberg J and Sj\"oqvist E 2005 {\it Phys. Rev. A} {\bf 72} 022328
\bibitem{paladino08} Paladino E, Sassetti M, Falci G and Weiss U 2008 {\it Phys. Rev. B} 77 041303


\bibitem{whitney03} Whitney R S and Gefen Y 2003 {\it Phys. Rev. Lett.} {\bf 90} 190402
\bibitem{whitney05} Whitney R S, Makhlin Y, Shnirman A and Gefen Y 2005 {\it Phys. Rev. Lett.} {\bf 94} 070407
\bibitem{whitney10} Whitney R S 2010 {\it Phys. Rev. A} {\bf 81} 032108

\bibitem{dechiara03} De Chiara G and Palma G M 2003 {\it Phys. Rev. Lett.} {\bf 91} 090404
\bibitem{solinas04} Solinas P, Zanardi P and  Zangh\`i N 2004 {\it Phys. Rev. A} {\bf 70} 042316
\bibitem{zhu05} Zhu S L and Zanardi P 2005 {\it Phys. Rev. A} {\bf 72} 020301(R)

\bibitem{kult06} Kult D,  Aberg J and Sj\"oqvist E 2006 {\it Phys. Rev. A} {\bf 74} 022106

\end{thebibliography}
\end{document}